\documentstyle[epsfig]{aipproc}

\begin{document}

\ifx\href\undefined
\else\errmessage{Don't use HyperTeX!}
\fi

\hbox{\rightline{\vbox{\hbox{CALT-68-2105}\hbox{hep-ph/9703269}}}}

\title{Connections Between Inclusive and \\
  Exclusive Semileptonic $B$ Decay%
\footnote{Talk given at the Latin American Symposium on High Energy Physics,
Merida, Mexico, 1996. }
}

\author{Mark B.\ Wise} 

\address{California Institute of Technology, 
  Pasadena, CA  91125, USA\thanks{Work supported in part by 
  the U.S.\ Dept.\ of Energy under Grant No.\ DE-FG03-92-ER40701.} }

\maketitle

\begin{abstract}
$B$ decay sum rules relate exclusive $B$ semileptonic decay matrix elements to
forward $B$-meson matrix elements of operators in HQET.  At leading order the
operators that occur are the $b$-quark kinetic energy $\lambda_1$ and
chromomagnetic energy $\lambda_2$.  The latter is determined by the measured
$B^*-B$ mass splitting.  The derivation of these sum rules is reviewed and
perturbative QCD corrections are discussed.  A determination of $\lambda_1$ and
the energy of the light degrees of freedom in a $B$-meson, $\bar\Lambda$, from
semileptonic $B$ decay data is presented.  Future prospects for improving these
sum rules are discussed.

\end{abstract}

\section{Introduction}

In this lecture I review some connections between inclusive and exclusive
semileptonic $B$-meson decays.  These arise from sum rules that relate the form
factors for exclusive semileptonic decays to nonperturbative QCD matrix
elements that occur in the inclusive semileptonic decay rate.  Sum rules that
relate inclusive $B$ transitions to a sum over exclusive states were first
derived by Bjorken~\cite{bjorken,isgur1} and Voloshin~\cite{voloshin1}.  Then a
general framework for $B$-decay sum rules was presented by Bigi, et al.\
in~\cite{bigi1}.  Since this very important work, there has been a considerable
amount of theoretical activity in the area of $B$ decay sum
rules 
[5--12].

For inclusive decays it is possible using the operator product expansion and a
transition to the heavy quark effective theory (HQET) \cite{eft} to show that at
leading order in $\Lambda_{\rm QCD}/m_b$ the $B$ semileptonic decay rate is
equal to the $b$-quark decay rate~\cite{chay,voloshin2}.  There are no
nonperturbative corrections to this at order $\Lambda_{\rm
QCD}/m_b$~\cite{chay}.  The first corrections arise at order $\Lambda_{\rm
QCD}^2/m_b^2$, and are characterized by matrix elements~\cite{bigi3} that are
related to the $b$-quark kinetic energy
\begin{equation}\label{1}
\lambda_1 = {1\over 2m_B} \langle B(v) |\bar h_v^{(b)} 
  (i D)^2 h_v^{(b)} | B(v)\rangle,
\end{equation}
and the color magnetic energy
\begin{equation}\label{2}
\lambda_2 = {1\over 6m_B} \langle B(v) | \bar h_v^{(b)} 
  {g\over 2} \sigma_{\mu\nu} G^{\mu\nu} h_v^{(b)} | B(v)\rangle.
\end{equation}
The parameters $\lambda_1$ and $\lambda_2$ are independent of the heavy quark 
mass and occur in the formulas for the $B$, $B^*$, $D$, and $D^*$ meson masses:
\begin{eqnarray}\label{3}
m_B &=& m_b + \bar\Lambda - (\lambda_1 + 3 \lambda_2) /2m_b, \nonumber \\*
m_{B^*} &=& m_b + \bar\Lambda - (\lambda_1 - \lambda_2) /2m_b, \nonumber \\*
m_D &=& m_c + \bar\Lambda - (\lambda_1 + 3 \lambda_2) /2m_c, \nonumber \\*
m_{D^*} &=& m_c + \bar\Lambda - (\lambda_1 - \lambda_2) /2m_c.
\end{eqnarray}
The measured $B^* - B$ mass splitting $(46 \pm 0.6)\,$MeV implies that
$\lambda_2 = 0.12\,{\rm GeV}^2$.  The quantity $\bar\Lambda$ represents the
energy of the light degrees of freedom for the ground state
$s_\ell^{\pi_{\ell}} = {1\over 2}^-$ multiplet  in the $m_{b,c} \rightarrow
\infty$ limit.  Note that in the average masses $\bar m_B = (m_B + 3m_{B^*})/4$
and $\bar m_D = (m_D + 3 m_{D^*})/4$ the parameter $\lambda_2$ cancels out.

The leading order prediction of the operator product expansion for the $B$
semileptonic decay rate involves quark masses, which are not known
experimentally.  What is measured are the hadron masses.  It is possible using
eq.~(\ref{3}) to express the quark masses, $m_b$ and $m_c$, in terms of the
hadron masses, $\bar m_B$ and $\bar m_D$, and the parameters $\lambda_1$ and
$\bar\Lambda$.  When this is done the semileptonic $B$-meson decay rate depends
on the unknown parameters $\lambda_1$ and $\bar\Lambda$ that are of order
$\Lambda^2_{\rm QCD}$ and $\Lambda_{\rm QCD}$ respectively.  In this way of
looking at the predicted decay rate there are contributions of order
$\Lambda_{\rm QCD}/m_{c,b}$, but they are given in terms of the single
parameter $\bar\Lambda$.

The form factors for semileptonic $B \rightarrow D^{(*)} e\bar\nu_e$ 
decay are defined by
\begin{eqnarray}\label{4}
{\langle D(v')|V^\mu|B(v)\rangle\over\sqrt{m_B m_D}} &=& 
  h_+ (w) (v + v')^\mu + h_- (w) (v - v')^\mu, \nonumber\\*
{\langle D^* (v',\epsilon)|V^\mu| B (v)\rangle\over \sqrt{m_B m_{D^*}}} &=& 
  ih_V (w) \varepsilon^{\mu\nu\alpha\beta} \epsilon_\nu^* v'_\alpha v_\beta, 
  \nonumber\\*
{\langle D^* (v',\epsilon)| A^\mu| B(v)\rangle\over\sqrt{m_B m_{D^*}}} &=& 
  h_{A_{1}} (w) (w + 1) \epsilon^{*\mu} 
  - h_{A_{2}} (w) (\epsilon^* \cdot v) v^\mu \nonumber\\*
&-& h_{A_{3}} (w) (\epsilon^* \cdot v) v^{\prime\mu}.
\end{eqnarray}
Here $V^\mu = \bar c \gamma^\mu b$ and $A^\mu = \bar c \gamma^\mu \gamma_5 b$
are the vector and axial vector currents.  The four-velocities of the initial
and final states are denoted by $v$ and $v'$ respectively.  The dot product of
these four-velocities is $w = v \cdot v'$ and at the zero recoil point, where
$v = v'$, $w = 1$.  Up to corrections suppressed by powers of
$\alpha_s(m_{c,b})$ and $\Lambda_{\rm QCD}/m_{c,b}$, $h_-(w) = h_{A_2}(w) = 0$
and $h_+(w) = h_V(w) = h_{A_1}(w) = h_{A_3}(w) = \xi(w)$, where the Isgur--Wise
function~\cite{isgur2} $\xi$ is evaluated at a subtraction point around
$m_{c,b}$.  The differential decay rates are
\begin{eqnarray}\label{5}
{{\rm d}\Gamma(B \rightarrow D^* \ell \bar\nu_e) \over {\rm d}w} &=& 
  {G_F^2 m_B^5\over 48\pi^3} r^{*3} (1 - r^*)^2 (w^2 - 1)^{1/2} (w + 1)^2
  \nonumber\\*
&\times& \left[1 + {4 w\over w + 1} {1 - 2 w r^* + r^{*2}\over 
  (1 - r^*)^2}\right] |V_{cb}|^2 | {\cal F}_{B\rightarrow D^{*}} (w)|^2, 
  \nonumber\\*
{{\rm d}\Gamma(B \rightarrow D \ell \bar\nu_e) \over {\rm d}w} &=& 
  {G_F^2 m_B^5\over 48\pi^3} r^3 (1 + r)^2 (w^2 - 1)^{3/2} 
  |V_{cb}|^2 |{\cal F}_{B\rightarrow D} (w)|^2,
\end{eqnarray}
where $r^{(*)}=m_{D^{(*)}}/m_B$.  
The functions ${\cal F}_{B \rightarrow D^{*}}$ and ${\cal F}_{B \rightarrow D}$ 
are given in terms of the form factors of the vector and axial vector currents
defined in eq.~(\ref{4}) as
\begin{eqnarray}
|{\cal F}_{B\rightarrow D^{*}} (w)|^2 &=& 
  \left[1 + {4w\over w + 1} {1 - 2 w r^* + r^{*2}\over (1 - r^*)^2}\right]^{-1}
  \nonumber\\*
&\times& \Bigg\{ {1 - 2 w r^* + r^{*2}\over (1 - r^*)^2}
 2 \left[ h_{A_{1}}^2 (w) + {w - 1\over w + 1} h_V^2 (w)\right] \nonumber\\*
&& + \Bigg[ h_{A_{1}} (w) + {w - 1\over 1 - r^*} \Bigg(h_{A_{1}} (w) 
  - h_{A_{3}} (w) - r^* h_{A_{2}} (w) \Bigg) \Bigg]^2 \Bigg\} , \nonumber\\*
{\cal F}_{B \rightarrow D} (w) &=& h_+ (w) - {1 - r\over 1 + r} h _- (w). 
\end{eqnarray}

Note that ${\cal F}_{B \rightarrow D^{*}}(1) = h_{A_{1}}(1)$ and due to 
Luke's theorem~\cite{luke}
\begin{equation}\label{7}
  h_{A_{1}} (1) = \eta_A + {\cal O} (\Lambda_{\rm QCD}^2/m_{cb}^2) .
\end{equation}
For the $B \rightarrow D$ case ${\cal F}_{B \rightarrow D} (1) = \eta_V + {\cal
O} (\Lambda_{\rm QCD}/m_{c,b})$.  The quantities $\eta_A$ and $\eta_V$ relate
the axial and vector currents in the full theory of QCD to those in HQET at
zero recoil.  Including corrections of order $\alpha_s^2
\beta_0$~\cite{voloshin3,neubert1}
\begin{eqnarray}\label{8}
\eta_A = 1 &-& {\alpha_s (\sqrt{m_b m_c})\over\pi} 
  \left({1 + z\over 1 - z} \ln z + {8\over 3}\right) \nonumber\\*
&-& {\alpha_s^2 (\sqrt{m_b m_c})\over \pi^2} \beta_0 {5\over 24} 
  \left({1 + z\over 1 - z} \ln z + {44\over 15}\right),
\end{eqnarray}
where $z = m_c/m_b$ and $\beta_0 = 11 - 2N_f/3$ is the 1-loop beta function. 
In eq.~(\ref{8}) and hereafter dimensional regularization with $\overline{\rm
MS}$ subtraction is used.  The full order $\alpha_s^2$ expression for $\eta_A$
is known~\cite{czarnecki} and the $\alpha_s^2 \beta_0$ part presented in
eq.~(\ref{8}) dominates it.

\newpage
\section{Sum Rules}

To derive the sum rules, we consider the time-ordered product
\begin{equation}\label{corrdef}
T_{\mu\nu} = {i\over 2m_B}\, \int{\rm d}^4x\, e^{-iq\cdot x}\,
  \langle B \,|\, T\{J_\mu^\dagger(x),J_\nu(0)\} | B\, \rangle \,,
\end{equation}
where $J_\mu$ is a $b\to c$ axial or vector current, the $B$ states are at
rest, $\vec q$ is fixed, and $q_0=m_B-E_{D^{(*)}}- \epsilon$.  Here
$E_{D^{(*)}}=\sqrt{m_{D^{(*)}}^2+|\vec q\,|^2}$ is the minimal possible energy
of the hadronic final states that can be created by the current $J_\mu$ at
fixed $|\vec q\,|$.  (We deal with cases where the lowest energy state is
either a $D$ or a $D^*$.)  With this definition of $\epsilon$, $T_{\mu\nu}$ has
a cut in in the complex $\epsilon$ plane that lies along $0 < \epsilon <
\infty$, corresponding to physical intermediate states with a charm quark.   At
the same value of $|\vec q|$ the cut at the parton level lies in the smaller
region $\epsilon > \bar\Lambda (w - 1)/w +{\cal O}(\Lambda^2_{\rm
QCD}/m_{c,b}^2)$, were $\vec q^2 = m_c^2 (w^2 - 1)$.  $T_{\mu\nu}$ has another
cut corresponding to physical states with two $b$-quarks and a $\bar c$ quark
that lies along $-\infty < \epsilon < -2 E_{D^{(*)}}$.  To separate out
specific hadronic form factors, one contracts the currents in
eq.~(\ref{corrdef}) with a suitably chosen four-vector $a$.  Inserting a
complete set of states between the currents yields 
\begin{equation}
a^{*\mu}\, T_{\mu\nu}(\epsilon)\, a^\nu = {1\over2m_B}\, 
  \sum_X\, (2\pi)^3\, \delta^3(\vec q+\vec p_X)\,
  {\langle B| J^\dagger\cdot a^* |X\rangle \langle X| J\cdot a |B\rangle \over
  E_X-E_{D^{(*)}}-\epsilon} + \ldots \,,
\end{equation}
where the ellipses denote the contribution from the cut corresponding to two
$b$-quarks and a $\bar c$ quark.  The sum over $X$ includes the usual phase
space factors, ${\rm d}^3p/(2\pi)^3 2E_X$, for each particle in the state $X$.

\begin{figure}[t]
\centerline{\epsfysize=9truecm \epsfbox{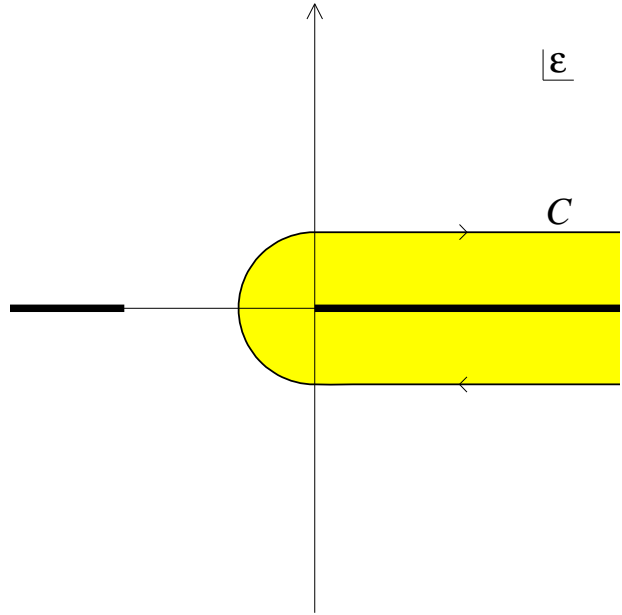}}
\caption[1]{The integration contour $C$ in the complex $\epsilon$ plane.
The cuts extend to ${\rm Re}\,\epsilon\to\pm\infty$. }
\end{figure}

While $T_{\mu\nu}(\epsilon)$ cannot be computed for arbitrary values of
$\epsilon$, its integrals with appropriate weight functions are calculable
using the operator product expansion and perturbative QCD.  Consider
integrating the product of a weight function $W_\Delta(\epsilon)$ with
$T_{\mu\nu}(\epsilon)$ along the contour $C$ surrounding the physical cut shown
in Fig.~1.  Assuming $W$ is analytic in the shaded region enclosed by this
contour, we get 
\begin{eqnarray}\label{zeroth}
{1\over2\pi i}\, && \int_C {\rm d}\epsilon\, W_\Delta(\epsilon)\, 
  [a^{*\mu}\, T_{\mu\nu}(\epsilon)\, a^\nu] \nonumber\\*
&& = \sum_X\, W_\Delta(E_X-E_{D^{(*)}})\, 
  (2\pi)^3\, \delta^3(\vec q + \vec p_X)\,
  {|\langle X| J\cdot a |B\rangle|^2\over2m_B}\, .
\end{eqnarray}
The weight function is assumed to be positive along the cut and to satisfy the
normalization condition $W_\Delta (0) = 1$.  Then $W_\Delta
\cdot |\langle X| J \cdot a|B\rangle|^2$ is positive for all states $X$, and
eq.~(\ref{zeroth}) implies an upper bound on the magnitude of form factors for
semileptonic $B$ decays to the ground states $D^{(*)}$.
\begin{equation}\label{12}
{|\langle D^{(*)} | J \cdot a | B\rangle |^2\over 4m_B E_{D^{(*)}}} 
  < {1\over 2\pi i} \int_C {\rm d}\epsilon\, W_\Delta (\epsilon) [a^{*\mu} 
  T_{\mu\nu} a^\nu].
\end{equation}
In eq.~(\ref{12}) a sum over $D^*$ polarizations is understood.
It is also possible to derive a lower bound if some model dependent 
assumptions concerning the spectrum of final states $X$ are made.

A possible set of weight functions is~\cite{kapustin1},
\begin{equation}\label{weightfn}
W_\Delta^{(n)}(\epsilon) = {\Delta^{2n}\over\epsilon^{2n}+\Delta^{2n}} \,,
  \qquad (n=2,3,\ldots) .
\end{equation}
They satisfy the following properties: ($i$) $W_\Delta$ is positive along the
cut so that every term in the sum over $X$ on the hadron side of the sum rule
is positive; ($ii$) $W_\Delta(0)=1$; ($iii$) $W_\Delta$ is flat near
$\epsilon=0$; and ($iv$) $W_\Delta$ falls off rapidly for $\epsilon>\Delta$. 
For values of $n$ of order unity all the poles of $W_\Delta^{(n)}$ lie at a
distance of order $\Delta$ away from the physical cut.  As $n\to\infty$,
$W_\Delta^{(n)}$ approaches $\theta(\Delta-\epsilon)$ for $\epsilon>0$, which
corresponds to summing over all final hadronic resonances up to excitation
energy $\Delta$ with equal weight.  In this limit the poles of $W_\Delta^{(n)}$
approach the cut, and the contour $C$ is forced to pinch the cut at $\epsilon =
\Delta$.  Then the evaluation of the contour integrals using
perturbative QCD relies on local duality at the scale $\Delta$.   In practice,
for $n > 3$ the results obtained are very close to those for $n = \infty$ and
for the remainder of this lecture I will only quote results obtained from the
weight $W_\Delta^{(\infty)} (\epsilon) = \theta (\Delta - \epsilon)$.

\section{Application of Sum Rules \\* at Zero Recoil}

The sum rule bound in eq.~(\ref{12}) is made explicit by using the operator
product expansion and perturbative QCD to evaluate the right-hand side.  The
most important kinematic point is the zero recoil point where $\vec q = 0$. 
Choosing $a$ to be a spatial vector $a = (0, \hat n)$ and averaging over
directions of the unit vector $\hat n$, we obtain for the axial vector current
\begin{eqnarray}\label{14}
|{\cal F}_{B\rightarrow D^{*}}(1)|^2 \leq \eta_A^2 &-& {\lambda_2\over m_c^2} 
  + \left( {\lambda_1 + 3\lambda_2\over 4}\right) \left({1\over m_c^2} 
  + {1\over m_b^2} + {2\over 3m_c m_b}\right) \nonumber\\*
&+& {\alpha_s (\Delta)\over \pi} X_{AA} (\Delta) 
  + {\alpha_s^2 (\Delta)\over\pi^2} \beta_0 Y_{AA} (\Delta),
\end{eqnarray}  
and for the vector current
\begin{eqnarray}\label{15}
0 &\leq& {\lambda_2\over m_c^2} - \left({\lambda_1 + 3\lambda_2\over 4} \right) 
  \left({1\over m_c^2} + {1\over m_b^2} - {2\over 3m_c m_b}\right) \nonumber\\*
&+& {\alpha_s(\Delta)\over\pi} X_{VV} (\Delta) 
  + {\alpha_s^2 (\Delta)\over\pi^2} \beta_0 Y_{VV} (\Delta).
\end{eqnarray}
Eqs.~(\ref{14}) and~(\ref{15}) include terms of order $\Lambda_{\rm
QCD}^2/m_c^2$ coming from dimension five operators in the operator product
expansion for the time ordered product of currents.  The coefficients of these
operators are evaluated at tree level.  Also included is the contribution from
the dimension 3 operator $\bar h_v^{(b)} \Gamma h_v^{(b)}$ evaluated to order
$\alpha_s^2 \beta_0$.  There are two distinct sources of perturbative QCD
corrections.  Those in $\eta_A$ correspond to a final state $X_c$ that at the
parton level is a single charm quark.  These terms are independent of $\Delta$
and come from matching of the axial vector current onto its HQET counterpart,
i.e., $A^\nu = \eta_A \bar h_v^{(b)} \gamma^\nu\gamma_5 h_v^{(c)}$.  The part
of the QCD correction involving $X_{AA}, Y_{AA}, X_{VV}$ and $Y_{VV}$, comes at
the parton level from states with a charm quark and a gluon or even more
partons.  These corrections depend on $\Delta$.  Since $\Delta$ is the cut off
on the invariant mass of the final hadronic states it seems most natural to
write these terms as a power series in $\alpha_s(\Delta)$.  If one used
$\alpha_s (\mu)$ with $\mu$ much different from $\Delta$ the coefficients
$Y_{AA}$ and $Y_{VV}$ would contain large logarithms of $\Delta/\mu$.  Analytic
expressions for the order $\alpha_s$ corrections are known~\cite{kapustin1} 
\begin{eqnarray}
X_{AA}(\Delta) &=& {\Delta (\Delta + 2m_c) 
  [2(\Delta + m_c)^2 - 2m_b^2 - (m_b + m_c)^2]\over 18 m_b^2 (\Delta + m_c)^2} 
  \nonumber \\*
&+& {3m_b^2 + 2m_bm_c - m_c^2\over 9m_b^2} 
  \ln \left({\Delta + m_c\over m_c}\right) , \label{16}\\*
X_{VV}(\Delta) &=& {\Delta (\Delta + 2m_c) 
  [2(\Delta + m_c)^2 - 2m_b^2 - (m_b - m_c)^2]\over 18 m_b^2 (\Delta + m_c)^2} 
  \nonumber \\*
&+& {3 m_b^2 - 2 m_b m_c - m_c^2\over 9m_b^2} 
  \ln \left({\Delta + m_c\over m_c}\right) . \label{17}
\end{eqnarray} 

For small $\Delta$, $X_{AA}$ and $X_{VV}$ are of order $\Delta^2/m^2_{c,b}$;
however, even when $\Delta = 1\,$GeV, terms higher order in $\Delta/m_{c,b}$
are important (the small $\Delta$ approximation to $X_{AA}$ and $X_{VV}$ was
calculated in Ref.~\cite{bigi1}).  The values of $Y_{AA}$ and $Y_{VV}$ have
been determined numerically~\cite{kapustin1}.  In Fig.~2, $X_{AA}$, $Y_{AA}$,
$X_{VV}$, and $Y_{VV}$ are plotted as functions of $\Delta$ in the region
$\Delta<2\,$GeV.  The values of $Y$ are quite close to $X$ in this region.  

\begin{figure}
\centerline{\epsfysize=7truecm \epsfbox{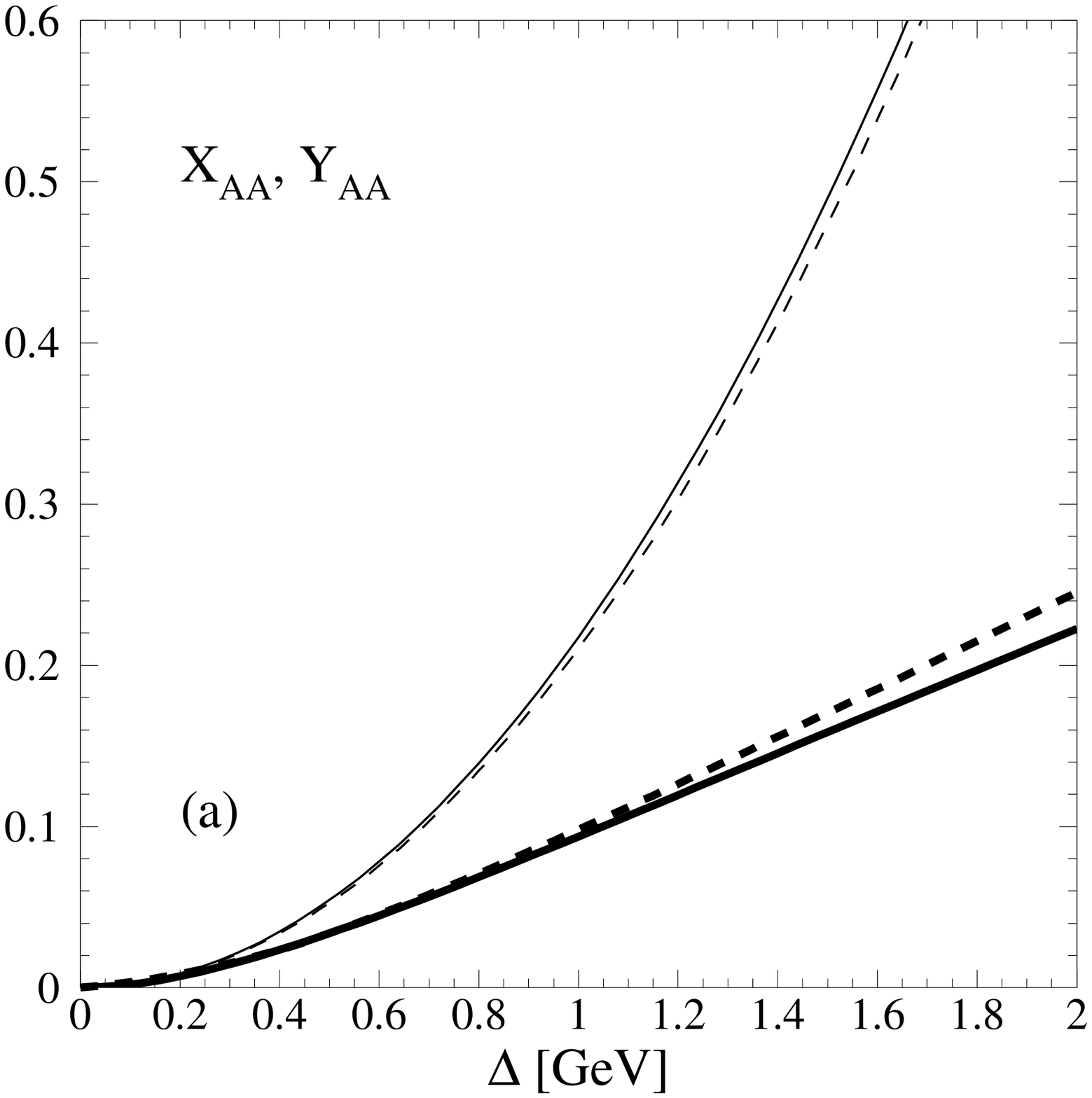} 
  \epsfysize=7truecm \epsfbox{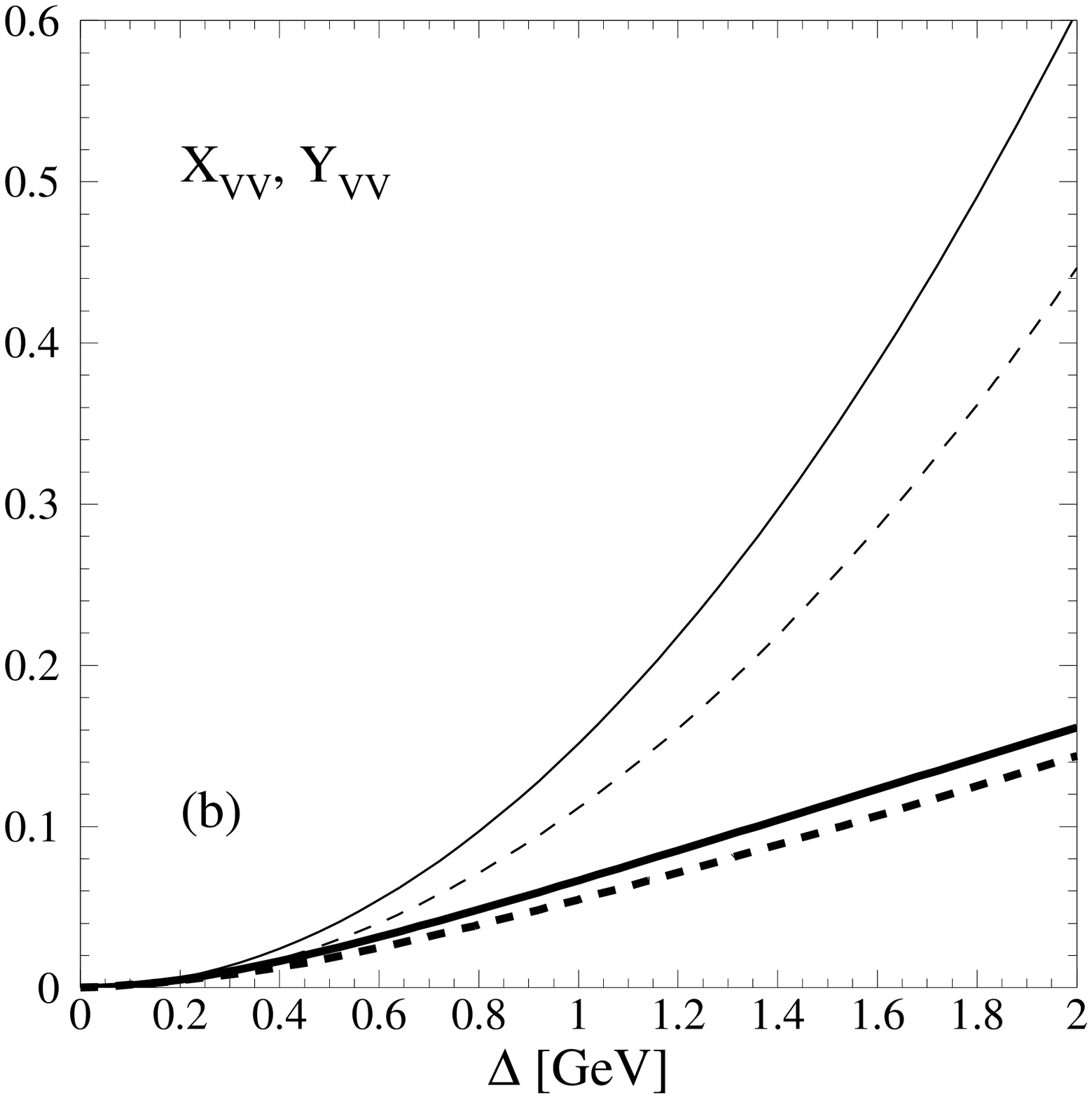}}
\caption[2]{$X(\Delta)$ and $Y(\Delta)$ for the a) axial, and b) vector 
coefficients.  Thick solid lines are $X$ while thick dashed lines are $Y$.  
The thin solid and dashed lines are $X$ and $Y$ to order $\Delta^2/m_{c,b}^2$. }
\end{figure}

The vector current sum rule bound in eq.~(\ref{15}) implies a bound on
$\lambda_1$.  This bound is strongest for $m_c \gg m_b \gg \Delta$.  In that
limit it becomes~\cite{bigi1,kapustin1}
\begin{equation}\label{18}
\lambda_1 \leq - 3\lambda_2 + {\alpha_s (\Delta)\over \pi} \Delta^2 
  \left({4\over 3}\right) + {\alpha_s^2 (\Delta)\over \pi^2} 
  \beta_0\Delta^2 \left({13\over 9} - {2\ln 2\over 3}\right).
\end{equation}

The parameter $\Delta$ must be chosen large enough that perturbative QCD is
meaningful.  However the bounds on $\lambda_1$ and $|{\cal F}_{B \rightarrow
D^*}|^2$ become stronger the smaller the value of $\Delta$.  The smallest value
of $\Delta$ for which one can imagine using perturbative QCD is $1\,$GeV.  
Using $\Delta = 1\,$GeV, $\alpha_s (1 {\rm GeV}) = 0.45$, 
$\lambda_2 = 0.12\,{\rm GeV}^2$, eq.~(\ref{18}) implies 
\begin{equation}\label{19}
  \lambda_1 \leq (- 0.36 + 0.19 + 0.20)\, {\rm GeV}^2.
\end{equation}
The three terms on the right-hand side of eq.~(\ref{19}) correspond
respectively to the contribution of $\lambda_2$, the perturbative part of order
$\alpha_s (\Delta) \Delta^2/\pi$, and the perturbative part of order $[\alpha_s
(\Delta)/ \pi]^2 \Delta^2$.  Notice that with $\Delta = 1\,$GeV the
$\alpha_s^2$ term is as large as the order $\alpha_s$ term.  It may be a
mistake to conclude from this that $\Delta = 1\,$GeV is too low for QCD
perturbation theory to be meaningful.  It has been conjectured that $\lambda_1$
has a renormalon ambiguity of order $\Lambda_{\rm QCD}^2$ (one does not see
this from the usual sum of bubble graphs)~\cite{neubert2}.  Even though the
renormalon ambiguity arises from large orders of perturbation theory, it is
possible that the bad behavior of the first few terms in the perturbative
series presented in eq.~(\ref{18}) is a reflection of this uncertainty.

In this lecture the matrix element $\lambda_1$ is defined using dimensional
regularization and $\overline{\rm MS}$ subtraction.  If $\lambda_1$ has a
renormalon ambiguity (of order $\Lambda_{\rm QCD}^2$), the perturbative QCD
series that relates it to a physical quantity, for example computed in lattice
QCD, is not Borel summable.  However, there is no evidence that this is a
serious problem.  Whenever $\lambda_1$ occurs in an expression for some
measurable quantity, e.g., the bound on $|{\cal F}_{B\rightarrow D^*}(1)|^2$,
there is another perturbative series that when combined with the series in
$\lambda_1$ (e.g., from matching onto lattice QCD) probably has no renormalon
ambiguity (of order $\Lambda^2_{\rm QCD}$)\cite{beneke}.

Next consider the bound on $|{\cal F}_{B\rightarrow D^*}(1)|^2$ in
eq.~(\ref{14}).  We can eliminate $\lambda_1$ from it by combining~(\ref{14})
and~(\ref{15}).  This gives
\begin{eqnarray}\label{20}
|{\cal F}_{B\rightarrow D^*}(1)|^2 &\leq& \eta_A^2 - {\lambda_{2}\over m_c^2}
  + {\alpha_s (\Delta)\over \pi} \left[X_{AA} (\Delta) 
  + {1\over 3}\! \left({\Delta^2\over m_c^2} + {\Delta^2\over m_b^2} 
  + {2\Delta^2\over 3m_cm_b}\right)\right] \\*
&+& {\alpha_s^2(\Delta)\over\pi^2} \beta_0 \left[Y_{AA} (\Delta) 
  + \left({13\over 36} - {\ln 2\over 6}\right) \left({\Delta^2\over m_c^2} 
  + {\Delta^2\over m_b^2} + {2\Delta^2\over 3m_cm_b}\right)\right] . \nonumber
\end{eqnarray}
Neglecting the nonperturbative correction factor of $-\lambda_2/m_c^2$ and 
again using $\Delta = 1\,$GeV, the above bound is
\begin{eqnarray}\label{21}
|{\cal F}_{B\rightarrow D^{*}}(1)|^2 \leq 1 &-& 0.074 - 0.020 \nonumber \\*
&+& 0.044 + 0.046 \nonumber \\*
= 1 &-& 0.030 + 0.026.
\end{eqnarray}
Here we used $m_c = 1.4\,$GeV, $m_b = 4.8\,$GeV, $\alpha_s (\sqrt{m_cm_b}) =
0.28$, $\alpha_s(1\,{\rm GeV}) = 0.45$, and $\beta_0 = 9$. The first row is the
perturbative expansion of $\eta_A^2$, the second row is the order $\alpha_s
(\Delta)$ and order $\alpha_s (\Delta)^2$ terms, and the third row sums the
columns.  There is a renormalon ambiguity of order $\Lambda_{\rm
QCD}^2/m_{c,b}^2$ that cancels between the perturbative series for $\eta_A^2$
and the series in $\alpha_s(\Delta)$.  This bound on the physical quantity
$|{\cal F}_{B\rightarrow D^{*}}(1)|^2$ is not very strong (even when the factor
of $-\lambda_2/m_c^2$ is included), and furthermore the third row of
eq.~(\ref{21}) seems to indicate that with $\Delta = 1\,$GeV QCD perturbation
theory is not very well behaved.  However, this does not mean that the sum rule
for $|{\cal F}_{B\rightarrow D^{*}}(1)|^2$ in eq.~(\ref{14}) is not useful. 
Consider the perturbative part of eq.~(\ref{14}), neglecting for now both the
terms of order $\lambda_1/m^2_{c,b}$ and $\lambda_2/m^2_{c,b}$.  Numerically,
with $\Delta = 1\,$GeV, this gives 
\begin{eqnarray}\label{22}
|{\cal F}_{B\rightarrow D^{*}}(1)|^2 \leq 1 &-& 0.074 - 0.020 \nonumber \\*
&+& 0.013 + 0.017 \nonumber\\*
= 1 &-& 0.061 - 0.003.
\end{eqnarray}
Again, the first row is the perturbative expansion of $\eta_A^2$ and the second
row are the terms of order $\alpha_s (\Delta)$ and $\alpha_s^2 (\Delta)$.  The
third row of eq.~(\ref{22}) does not indicate that there is any breakdown of
QCD perturbation theory.  If $\lambda_1$ can be determined experimentally from,
for example, the electron spectrum in inclusive semileptonic $B$-decay then the
sum rule in eq.~(\ref{14}) may lead to an important constraint.  For example,
with $\Delta = 1\,$GeV and $\lambda_1 = - 0.2\,{\rm GeV}^2$, eq.~(\ref{14})
implies the bound
\begin{equation}
  |F_{B\rightarrow D^{*}}(1)|^2 \leq 0.90 ,
\end{equation}
which is smaller than $\eta_A^2 = 0.91$.

\section{The Lepton Energy Spectrum and \\
  the Parameters $\lambda_1$, $\bar\Lambda$}

The CLEO collaboration has measured the lepton energy spectrum for inclusive $B
\to X \ell\bar \nu_e$ decay, both demanding only one charged lepton (single
tagged data) and two charged leptons (double tagged
sample)~\cite{bartlett,wang}.  In the double tagged sample the charge of the
high momentum lepton determines whether the other lepton comes from a
semileptonic $B$ decay (primary lepton) or the semileptonic decay of a
$D$-meson (secondary lepton).  The single tagged sample is presented in
$50\,$MeV bins while the double tagged data is presented in $100\,$MeV bins. 
The single tagged sample has much higher statistics, but is significantly
contaminated by secondaries below $E_\ell = 1.5\,$GeV.

The operator product expansion for semileptonic $B$ decay does not reproduce
the physical lepton spectrum point by point near the maximal electron energy. 
Near the endpoint, comparison of theory with data can only be made after
smearing or integrating over a large enough region.  The minimal size of this
region has been estimated to be about $500\,$MeV.  As was mentioned in the
introduction, the theoretical prediction for the lepton energy spectrum depends
on $\lambda_1$ and $\bar\Lambda$, so we can try to use the data to determine
these quantities.  We want to consider observables sensitive to $\bar\Lambda$
and $\lambda_1$, but we also want deviations from the $b$-quark decay rate to
be small enough so that contributions from even higher dimension operators in
the operator product expansion are small.  Ref.~\cite{gremm1} uses $R_1$ and
$R_2$, where 
\begin{equation}\label{23}
R_1 = {\int_{1.5\,{\rm GeV}} ({\rm d}\Gamma/{\rm d}E_\ell)E_\ell\, {\rm d}E_\ell
  \over \int_{1.5\,{\rm GeV}} ({\rm d}\Gamma/{\rm d}E_\ell)\, {\rm d}E_\ell} ,
\end{equation}
and
\begin{equation}\label{24}
R_2 = {\int_{1.7\,{\rm GeV}} ({\rm d}\Gamma/{\rm d}E_\ell)\, {\rm d}E_\ell\over
  \int_{1.5\,{\rm GeV}} ({\rm d}\Gamma/{\rm d}E_\ell)\, {\rm d}E_\ell}. 
\end{equation}
Here $E_\ell$ denotes the lepton energy.  The variable $R_1$ has dimensions of
mass and values for it will be given in GeV.  Ratios are considered so
that $|V_{cb}|$ cancels out.  Before comparing the experimental data with
theoretical predictions derived from the operator product
expansion and QCD perturbation theory, it is necessary to include
electromagnetic corrections and effects of the boost to the laboratory frame. 
This gives
\begin{eqnarray}\label{25}
R_1 &=& 1.8059 - 0.309 \left({\bar\Lambda\over \bar m_B}\right) 
  - 0.35 \left({\bar\Lambda\over\bar m_B}\right)^2
  - 2.32 \left({\lambda_1\over\bar m_B^2} \right) 
  - 3.96 \left({\lambda_2\over\bar m_B^2}\right) \nonumber\\*
&-& {\alpha_s\over \pi} \left( 0.035 + 0.07 {\bar\Lambda\over\bar m_B}\right)
  + \left|{V_{ub}\over V_{cb}}\right|^2 
  \left( 1.33 - 10.3 {\bar\Lambda\over\bar m_B}\right) \nonumber\\*
&-& \left(0.0041 - 0.004  {\bar\Lambda\over\bar m_B}\right) 
  + \left(0.0062 + 0.002 {\bar\Lambda\over\bar m_B}\right) ,
\end{eqnarray}
and
\begin{eqnarray}\label{26}
R_2 &=& 0.6581 - 0.315 \left({\bar\Lambda\over\bar m_B}\right) 
  - 0.68 \left({\bar\Lambda\over\bar m_B}\right)^2 
  - 1.65 \left({\lambda_1\over\bar m_B^2}\right)
  - 4.94 \left({\lambda_2\over\bar m_B^2}\right) \nonumber\\*
&-& {\alpha_s\over\pi} \left( 0.039 + 0.18 {\bar\Lambda\over\bar m_B} \right) 
  + \left|{V_{ub}\over V_{cb}}\right|^2 
  \left(0.87 - 3.8 {\bar\Lambda\over\bar m_B}\right) \nonumber\\*
&-& \left(0.0073 + 0.005 {\bar\Lambda\over\bar m_B}\right) 
  + \left( 0.0021 + 0.003  {\bar\Lambda\over\bar m_B}\right) .
\end{eqnarray}
In eqs.~(\ref{25}) and (\ref{26}) the charm and bottom quark masses have been
expressed in terms of $\bar m_B$, $\bar m_D$, $\bar\Lambda$, $\lambda_1$, and
$\lambda_2$ using eq.~(\ref{3}), which is why $\bar\Lambda$ occurs in these
formulas.  The last two terms in eqs.~(\ref{25}) and (\ref{26}) are from
electromagnetic radiative corrections and from the boost to the laboratory
frame respectively.  The experimental values for $R_1$ and $R_2$ are $R_1 =
1.7831\,$GeV and $R_2 = 0.6159$.  These were obtained from the single tagged
data with a correction for the secondary leptons obtained from the double
tagged sample.  For $R_1$ this correction is $0.0001\,$GeV and for $R_2$ it is
$0.0051$.  Comparing experiment with eqs.~(\ref{25}) and~(\ref{26}) gives the
central values $\bar\Lambda = 0.39 \pm 0.11\,$GeV and $\lambda_1 = - 0.19 \pm
0.10\,{\rm GeV}^2$.  Fig.~3 shows the one sigma bands on the allowed values of
$\bar\Lambda$ and $\lambda_1$ from $R_1$ and $R_2$.  The narrower band
corresponds to the $R_1$ constraint.  The shaded ellipse is the one sigma
allowed region for $\bar\Lambda$ and $\lambda_1$ including correlations between
$R_1$ and $R_2$.  The errors included in this analysis are just the statistical
ones.  An analysis of systematic errors has not been performed.  However, they
are only weakly energy dependent for $E_\ell \geq 1.5\,$GeV and it is hoped
that for $R_{1,2}$ systematic errors are smaller than the statistical ones. 
Note that the bands from $R_1$ and $R_2$ are almost parallel, so even small
corrections can significantly change the central values for $\bar\Lambda$ and
$\lambda_1$ obtained from Fig.~3.

\begin{figure}[tb]  
\centerline{\epsfxsize=9truecm \epsfbox{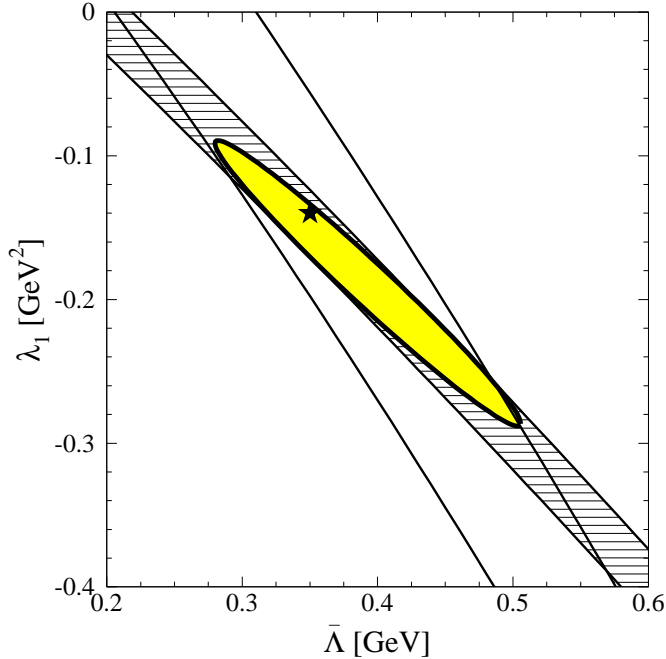}}
\caption[3]{Allowed regions in the $\bar\Lambda-\lambda_1$ plane
for $R_1$ and $R_2$.  The bands represent the $1\sigma$ statistical 
errors, while the ellipse is the allowed region taking correlations into 
account.  The star denotes where the order $\Lambda_{\rm QCD}^3/\bar m_B^3$ 
corrections discussed in the text would shift the center of the ellipse. }
\end{figure}

One such set of corrections comes from higher dimension operator in the
operator product expansion.  At order $\Lambda_{\rm QCD}^3/\bar m_B^3$ new
terms occur characterized by two matrix elements $\rho_1$ and $\rho_2$ and two
time ordered products.  $\rho_1$ can be estimated by factorization, $\rho_1 =
(2\pi \alpha_s/9)m_B f_B^2 \approx (300\,{\rm MeV})^3$, and $\rho_2$ is
expected to be small~\cite{bigi1}.  Neglecting $\rho_2$ and the time ordered
products gives the following order $\Lambda_{\rm QCD}^3/\bar m_B^3$ corrections
to $R_1$ and $R_2$,
\begin{eqnarray}
\delta R_1 &=& - (0.4 \bar\Lambda^3 + 5.7 \bar\Lambda\lambda_1 
  + 6.8 \bar\Lambda\lambda_2 + 7.7 \rho_1)/\bar m_B^3 , \nonumber \\*
\delta R_2 &=& - (1.5 \bar\Lambda^3 + 7.1 \bar\Lambda\lambda_1 
  + 17.5 \bar\Lambda\lambda_2 + 1.8 \rho_1)/\bar m_B^3 .
\end{eqnarray}
Including these corrections shifts the central values for $\bar\Lambda$ and
$\lambda_1$ to the location $\bar\Lambda = 0.35\,$GeV and $\lambda_1 = -
0.15\,{\rm GeV}^2$ marked by the star in Fig.~3.  For a more complete
discussion of the order $\Lambda_{\rm QCD}^3/\bar m_B^3$ corrections, see
Ref.~\cite{AM}.

The bands in Fig.~3 were determined using $|V_{ub}/V_{cb}| = 0.08$.  This value
is model dependent.  If $|V_{ub}/V_{cb}| = 0.10$ is used then the central
values shift to $\bar\Lambda = 0.42\,$GeV and $\lambda_1 = - 0.19\,{\rm
GeV}^2$.  In Fig.~3, $\alpha_s = 0.22$ was used corresponding to a subtraction
point near $m_b$.  With $\alpha_s = 0.35$ the central values shift to
$\bar\Lambda = 0.36\,$GeV and $\lambda_1 = - 0.18\,{\rm GeV}^2$.

Theoretical uncertainty in this determination of $\bar\Lambda$ and $\lambda_1$
originate from the reliability of quark hadron duality at the limits of
integration defining $R_{1,2}$, order $\Lambda_{\rm QCD}^3/\bar m_B^3$
corrections, and higher order perturbative QCD corrections.  Recently the order
$\alpha_s^2 \beta_0$ terms in $R_1$ and $R_2$ have been computed~\cite{gremm2}.
They give corrections $\delta R_1 = -0.082 \alpha_s^2 \beta_0/\pi^2$ and
$\delta R_2 = -0.098 \alpha_s^2 \beta_0/\pi^2$, moving the central values to
$\bar\Lambda = 0.33\,$GeV and $\lambda_1 = - 0.17\,{\rm GeV}^2$.  Concerning
duality, note that the lower limits on the lepton energy $E_\ell \geq 1.5\,$GeV
and $E_\ell \geq 1.7\,$GeV used in $R_{1,2}$ correspond to summing over
hadronic states $X$ with masses less than $3.6\,$GeV and $3.3\,$GeV
respectively.   Changing the lower limit in the numerator of $R_2$ to
$1.8\,$GeV leads to central values $\bar\Lambda = 0.47\,$GeV and $\lambda_1 = -
0.26\,{\rm GeV}^2$.  The plot in Fig.~3 uses electron data only.  Using the
muon sample instead gives compatible central values of $\bar\Lambda =
0.43\,$GeV and $\lambda_1 = - 0.21\,{\rm GeV}^2$.

It would be nice to have another constraint on $\lambda_1$ and $\bar\Lambda$
that would be less parallel than $R_1$ and $R_2$ are.  This can be provided by
the photon spectrum in inclusive $B \rightarrow X_s \gamma$
decay~\cite{kapustin2} which, when the data improves, will give a band almost
parallel to the $\lambda_1$ axis of Fig.~3.  Lattice QCD can also be used to
determine $\lambda_1$ and $\bar\Lambda$, although for $\lambda_1$ there are
serious difficulties coming from mixing with the lower dimension operator $\bar
h_v^{(b)} h_v^{(b)}$ \cite{gimenez}.  This mixing does not occur in the
continuum if dimensional regularization with $\overline{\rm MS}$ subtraction is
used.

\section{Concluding Remarks}

In this lecture I have reviewed the derivation of $B$-decay sum rules,
discussed the perturbative QCD corrections, and reviewed the status of the
determination of the nonperturbative QCD matrix element $\lambda_{1}$ that
occurs in the sum rules.  

If the contribution of the lowest lying excited states $X$ on the right-hand
side of eq.~(\ref{zeroth}) were known then this would imply a better bound on
the ground state matrix elements.  The lowest lying excited states are
nonresonant $D^{(*)} \pi$.  Their contribution, for low $D^{(*)}\pi$ invariant
mass, is calculable~\cite{bigi1} in terms of the one coupling constant, $g$, of
heavy hadron chiral perturbation theory~\cite{wise}.  This coupling also
determines the $D^*$ width, $\Gamma (D^{*+} \rightarrow D^0\pi^+) = (g^2/6\pi
f_\pi^2)|\vec p_\pi|^3$ (for the neutral pion mode there is an additional
factor of 1/2).  Unfortunately at the present time there is only a limit on the
$D^*$ width and hence an upper bound on $g$.  A measurement of the $D^*$ width
would give a direct determination of this coupling.  Then we would  know the
contribution of these nonresonant states to the sum rules.  (Determining $g$
from various $D^*$ and $D_s^*$ branching ratios is discussed in
Ref.~\cite{amundsen}).  Higher in mass is the $s_\ell^{\pi_{\ell}} = \frac32^+$
doublet of excited charmed mesons $D_1(2420)$ and $D_2^*(2460)$.  These states
are narrow with widths around $20\,$MeV.  A doublet with $s_\ell^{\pi_{\ell}} =
\frac12^+$ quantum numbers is expected to also exist, but these states are
thought to be quite broad (i.e., widths greater than $100\,$MeV).  Very
recently the contribution of these excited charmed mesons to the sum rules has
been discussed~\cite{leibovich}.

Perturbative QCD corrections to the sum rules have been calculated to order
$\alpha_s^2 \beta_0$.  It is important to improve this to a full order
$\alpha_s^2$ calculation.  Finally, it is interesting to note that $B$-decay
sum rules may also be important for the $b \rightarrow u$ transition, giving
valuable information on exclusive $B \rightarrow \pi \ell \bar\nu_\ell$ or $
B\rightarrow \rho \ell \bar\nu_\ell$ form factors~\cite{boyd1}.  At this time
perturbative corrections to these $b \rightarrow u$ transition sum rules have
not been computed.

\end{document}